\documentstyle[preprint,aps,eqsecnum]{revtex}

\begin{document}
\draft
\preprint{\vbox{
\hbox{CTP-TAMU-77/93}
\hbox{TRI-PP-94-2}
\hbox{hep-ph/9401228}
\hbox{December 1993}
}}

\title{
Lepton flavor changing processes and CP violation\\
in the 331 model
}

\author{James T. Liu}
\address{
Center for Theoretical Physics, Department of Physics\\
Texas A\&M University, College Station, TX 77843--4242
}
\author{Daniel Ng}
\address{
TRIUMF, 4004 Wesbrook Mall\\
Vancouver, B.C., V6T 2A3, Canada
}
\maketitle
\begin{abstract}
By extending the electroweak gauge group to $SU(3)_L\times U(1)_Y$, the 331
model incorporates dilepton gauge bosons $Y$ which do not respect individual
lepton family number.  We point out that, in addition to family diagonal
couplings such as $Y$--$e$--$e$ that change lepton family number by two units,
dileptons may also have family non-diagonal couplings such as
$Y$--$\mu$--$e$.  The latter coupling violates lepton family number by a
single unit and manifests itself via lepton flavor changing decays such as
$\mu\to3e$ and $\mu\to e\gamma$.  The family non-diagonal interaction can
be CP violating and typically generates extremely large leptonic electric
dipole moments.  We demonstrate a natural mechanism for eliminating both
single unit lepton flavor violation and large leptonic CP violation.
Although we focus on the 331 model, our results are applicable to other
dilepton models as well, including $SU(15)$ grand unification.
\end{abstract}
\pacs{PACS numbers: 12.15.Ff, 13.35.+s, 11.30.Er, 12.15.Cc}

\narrowtext

\section{Introduction}
While the Standard Model (SM) is extremely successful and is consistent with
known experimental data, it nevertheless leaves some questions unexplained.
Among these questions is the issue of why there are exactly three families
of quarks and leptons.  The 331 model gives a natural answer to this family
replication question and furthermore gives some indication as to why the top
quark is so heavy.

In the 331 model, the $SU(2)_L\times U(1)_Y$ electroweak gauge group of the
SM is extended to $SU(3)_L\times U(1)_X$ \cite{pleitez,frampton}.  Unlike
the SM, where anomalies cancel family by family, anomalies in the 331 model
only cancel when all {\it three} families are taken together.  This is
accomplished by choosing one of the families, which we take as the third
one, to transform differently under the 331 gauge group.  A different third
family conveniently allows a heavy top, but also introduces tree level
flavor changing neutral currents (FCNC).

Since the 331 model reduces to the standard electroweak theory, tree
level FCNC is restricted to interactions not present in the SM.  In
the gauge sector, only the new neutral gauge boson $Z'$ has a flavor changing
coupling to the ordinary quarks \cite{pleitez,daniel}.  Because the leptons
are treated democratically, they do not suffer FCNC (ignoring possible
flavor changing neutral Higgs interactions).  In the SM, the absence
of FCNC and massless neutrinos is sufficient to show that individual lepton
flavors are conserved.  While both conditions are true in the minimal 331
model, it turns out that lepton flavor is no longer conserved.  Lepton
flavor violation occurs through the interactions of the dilepton gauge
bosons $Y^+$ and $Y^{++}$ which both carry two units of lepton number.
Since dileptons do not carry lepton family information, only the total
lepton number, $L\equiv L_e+L_\mu+L_\tau$, is conserved (in the absence of
anomalies).

It is well known that dilepton interactions may violate individual lepton
family number by two, for instance in the process $e^-e^-\to Y^{--}\to
\mu^-\mu^-$, yielding spectacular signatures for dilepton models
\cite{frampmpl}.  However, little attention has been placed on the
possibility of single unit lepton flavor violation in these models.
Experimentally, the non-observation of such decays as $\mu\to3e$ and
$\mu\to e\gamma$ put strong constraints on $\Delta L_i=\pm1$ processes.  In
this paper, we examine the leptonic sector of the 331 model in detail and
study the dilepton contributions to lepton flavor violation.  While lepton
flavor violation universally occurs in the presence of massive neutrinos,
such contributions are often extremely small due to a GIM cancellation.  We
show that, even with massless neutrinos, the 331 model allows possibly large
lepton flavor violation mediated by dilepton exchange.

Unlike the SM, dilepton exchange may also contribute to large CP violation
in the leptonic sector.  This occurs because additional phases are present
in the mixing matrix describing the lepton couplings to the dilepton gauge
bosons.  These phases remain even with massless neutrinos, and cannot be
rotated away.  We examine the possibility of detecting such CP
violation by calculating the dilepton contributions to leptonic electric
dipole moments (EDM).  Our results show that dilepton mediated leptonic CP
violation may be extremely large, and is closely related to lepton flavor
violation.

Another source of CP violation in the 331 model is that coming from the
Higgs sector.  Since the minimal 331 model requires four Higgs multiplets,
there are many possibilities for Higgs sector CP violation.  In order to
examine such scenarios, we present a detailed discussion of the minimal 331
Higgs sector and show how it reduces to a three Higgs doublet SM with
additional $SU(2)_L$ singlet and triplet scalars carrying lepton number.
While a three Higgs doublet model gives a natural framework for
spontaneous CP violation \cite{weinberg,deshpande,branco}, we note that
both tree level flavor changing neutral Higgs (FCNH) \cite{lee,liuw}
and the additional singlet and triplet scalars \cite{zee} present additional
mechanisms for CP violation in the 331 model.

In order for the 331 model to be consistent with stringent experimental
bounds on lepton flavor violation and lepton EDMs, we find that the family
non-diagonal dilepton couplings must be very small.  We show that a natural
solution is to simply set them to zero (at least at tree level) which may be
accomplished by restricting the lepton Yukawa couplings by an appropriate
discrete symmetry.  An interesting feature of our analysis is that, while
the details are specific to the 331 model, the general results hold for any
model incorporating dilepton gauge bosons such as $SU(15)$ grand unification
\cite{su15fl,su15fk,su15fn,su15afn}.

In the next section we present a quick review of the 331 model and its
particle content.  In section 3, we examine the breaking of the 331 model
to the SM and show how CP violation may arise in the reduced Higgs sector.
In section 4, we show how $\Delta L_i=\pm1$ lepton flavor violation occurs
and study the related leptonic CP violation.  We present our
conclusions in section 5.  Details on the diagonalization of the charged
lepton mass matrix are given in an appendix.

\section{A review of the 331 model}

Construction of the 331 model was first presented in
Refs.~\cite{pleitez,frampton} and subsequently expanded upon in
Refs.~\cite{daniel,foot}.  In this section, we present a brief review of the
model.  Since the original papers have used a variety of different notations,
this review also serves to set up the conventions used in this paper.

\subsection{fermion representations}
Since each lepton family has three helicity states (assuming massless
neutrinos), they fall naturally into $SU(3)_L$ anti-triplets
\cite{fn:triplet}
\begin{equation}
\psi_i=\pmatrix{\ell_i^-\cr -\nu_i\cr \ell_i^+}_L\ ,
\label{eq:lepton}
\end{equation}
where $i=1,2,3$ is a family index.  We choose the standard embedding of
$SU(2)$ in $SU(3)$ (given by $T^a={1\over2}\lambda^a$ for triplets where
$\lambda^a$ are the usual Gell-Mann matrices)
so that the first two components of (\ref{eq:lepton})
corresponds to the ordinary electroweak doublet.  As a result, we find that
the hypercharge is given by $Y/2=\sqrt{3}T^8+X$ where leptons have vanishing
$X$ charge, $X=0$.  Our choice of hypercharge corresponds to twice the
average electric charge of $SU(2)_L$ representations, {\it i.e.}~$Q=T^3+Y/2$.
Thus each lepton family is in the $({\bf1},{\bf3}^*)_0$ representation of
$SU(3)_c \times SU(3)_L \times U(1)_X$.  A result of this embedding
is that there are {\it no} new leptons in the 331 model.

Note that upon reduction to $SU(2)$, both
$SU(3)$ triplets and anti-triplets decompose into a doublet and a singlet.
Since $SU(2)$ is pseudo-real, there is no distinction between these two
cases.  However, in order to get rid of some unimportant phases, we multiply
standard $SU(2)$ doublets by $i\tau^2=\left({\phantom{-}0\atop-1}\,
{1\atop0}\right)$ before
embedding them into $SU(3)$ anti-triplets.  This is the origin of the minus
sign in Eq.~(\ref{eq:lepton}).

While all three lepton families are treated identically, anomaly cancellation
requires that one of the three quark families transform differently from
the other two \cite{pleitez,frampton}.  In particular, cancelling the
pure $SU(3)_L$ anomaly requires the same number of triplets as anti-triplets.
Since there are three lepton anti-triplets and three quark colors,
we find that anomaly cancellation requires that two families of quarks
transform as triplets, $({\bf3},{\bf3})_{-1/3}$, whereas the third
transforms as an anti-triplet, $({\bf3},{\bf3}^*)_{2/3}$.  All left handed
anti-particles are put in as singlets in the usual manner,
$({\bf3}^*,{\bf1})_{-2/3,1/3,4/3}$ for the first two
families and $({\bf3}^*,{\bf1})_{-5/3,-2/3,1/3}$ for the third.  We will not
elaborate any further on the quarks.

\subsection{the gauge sector}
When the electroweak gauge group is extended to $SU(3)_L\times
U(1)_X$, we find 5 new gauge bosons beyond the SM.  We denote the $SU(3)_L$
gauge bosons by $W_\mu^a$ ($a=1\ldots8$) with $a=1,2,3$ forming the
$SU(2)_L$ subgroup of $SU(3)_L$.  The $U(1)_X$ gauge boson is given by
$X_\mu$.
We define the two gauge couplings, $g$ and $g_X$ according to
\begin{equation}
D_\mu=\partial_\mu-i gT^aW_\mu^a-ig_X {X\over\sqrt{6}}X_\mu\ ,
\label{eq:covar}
\end{equation}
with the conventional non-abelian normalization ${\rm Tr} T^aT^b={1\over2}
\delta^{ab}$ in the fundamental (triplet) representation.
The factor $1/\sqrt{6}$ was chosen \cite{frampton,daniel} so that for
triplets $X/\sqrt{6}\equiv T^9X$ with ${\rm Tr} T^9T^9={1\over2}$.

{}From above, we have found the hypercharge to be given by
$Y/2=\sqrt{3}T^8+X=\sqrt{3}T^8+\sqrt{6}T^9X$.  As a result, when 331 is
broken to the SM, we find the gauge matching conditions
\begin{equation}
{1\over g'^2}={3\over g^2}+{6\over g_X^2}\ ,
\label{eq:match}
\end{equation}
where the $U(1)_Y$ coupling constant $g'$ is given by $\tan\theta_W=g'/g$.
The consequences of this relation will be explored in the next section where
the reduction to the SM is carried out in more detail.

Since ${\bf8}_0\to
{\bf3}_0+{\bf2}_3+{\bf2}_{-3}+{\bf1}_0$ under $SU(3)_L\times U(1)_X\to
SU(2)_L\times U(1)_Y$, the new gauge bosons form a complex
$SU(2)_L$ doublet of dileptons, $(Y^{++},Y^+)$ with hypercharge $3$ and a
singlet, $W^8$.  This new $U(1)$ gauge boson $W^8$ mixes with the $U(1)_X$
gauge boson $X$ to give the hypercharge boson $B$ and a new $Z'$.

\subsection{Higgs fields}
At first glance, only two Higgs representations are necessary for symmetry
breaking, one to break 331 to the SM and the other to play the role of the
SM Higgs.  However, the Yukawa couplings are restricted by $SU(3)_L$ gauge
invariance.  In order to give realistic masses to all the particles, there
must be a minimum of four Higgs in the 331 model \cite{fn:alt}.
These four multiplets are the three triplets, $\Phi$, $\phi$ and $\phi'$
in representations $({\bf1},{\bf3})_1$, $({\bf1},{\bf3})_0$ and
$({\bf1},{\bf3})_{-1}$ respectively, and a sextet $({\bf1},{\bf6})_0$
denoted $H$.

$SU(3)_L\times U(1)_X$ is broken to $SU(2)_L\times U(1)_Y$ when $\Phi$
acquires a VEV, giving masses to the $Y$ and $Z'$ gauge bosons and the
new quarks.
At this stage of symmetry breaking, the other three Higgs fields
decompose into $SU(2)_L\times U(1)_Y$ representations as ${\bf3}_0\to
{\bf2}_1+{\bf1}_{-2}$, ${\bf3}_{-1}\to {\bf2}_{-1}+{\bf1}_{-4}$ and
${\bf6}_0\to {\bf3}_2+{\bf2}_{-1}+{\bf1}_{-4}$.  Taking this decomposition
into account, we may write the Higgs fields explicitly in terms of $SU(2)_L$
component fields as
\begin{equation}
\Phi=\pmatrix{\Phi_Y\cr\varphi^0}\qquad
\phi=\pmatrix{\Phi_1\cr\Delta^-}\qquad
\phi'=\pmatrix{\widetilde{\Phi}_2\cr\rho^{--}}\ ,
\label{eq:tripletH}
\end{equation}
and
\begin{equation}
H=\pmatrix{T&\widetilde{\Phi}_3/\sqrt{2}\cr
\widetilde{\Phi}_3^T/\sqrt{2}&\eta^{--}}\ .
\label{eq:sextetH}
\end{equation}
In the above, $\Phi_Y=(\Phi_Y^{++},\Phi_Y^+)$ is the Goldstone boson doublet
``eaten'' by the dileptons.  $\Phi_i=(\phi_i^+,\phi_i^0)$ $(i=1,2,3)$ are
three standard model Higgs doublets where
$\widetilde{\Phi}_i=i\tau^2\Phi_i^*$, and $T$ is an $SU(2)_L$ triplet,
\begin{equation}
T=\pmatrix{T^{++}&T^+/\sqrt{2}\cr T^+/\sqrt{2}&T^0}\ .
\end{equation}
As a result, the scalars give rise to a three Higgs doublet SM
with an additional $SU(2)_L$ triplet and charged singlets.

\subsection{lepton number assignment}
Because both the charged lepton and its anti-particle are in the same
multiplet, the assignment of lepton number is not entirely obvious.
Starting with $L(\ell^-)=L(\nu)=1$ and $L(\ell^+)=-1$, we find that the
dilepton doublet $(Y^{++},Y^+)$ carries lepton number $L=-2$.
Lepton numbers for the scalars may be assigned by inspection of the Yukawa
couplings.  We find that $\Phi_Y$ and $T$ carry lepton number $L=-2$ and
$\Delta^-$, $\rho^{--}$ and $\eta^{--}$ have $L=2$.  $\varphi^0$ and the SM
Higgs doublets carry no lepton number as expected.  This assignment is
consistent with the scalars giving rise to the longitudinal components of
the dilepton gauge bosons, even after $SU(2)_L$ breaking.

Given the above assignment of lepton number, the only place where it may be
explicitly violated is in the scalar potential.  This may be done either via
soft (dimension three) or hard (dimension four) terms.  In addition, the
triplet $T$ (with $L=-2$) has a neutral component which may acquire a VEV
and spontaneously break lepton number.  These possibilities may be
classified as follows:
\begin{itemize}
\item {\it no explicit $L$ violation and $\langle T\rangle=0$}:  This is the
minimal 331 model where total lepton number is conserved.  However, because
of the presence of dilepton gauge bosons, individual lepton family number
may be violated.  The parameters of the Higgs potential may be chosen so
that there is a stable minimum which maintains $\langle T\rangle=0$
\cite{foot,pleitez2}.
\item {\it no explicit $L$ violation but $\langle T\rangle\ne0$}:  In this
case, lepton number is spontaneously broken, thus leading to a triplet
Majoron model \cite{gelmini}.  This case is ruled out experimentally by $Z$
lineshape measurements.
\item {\it explicit $L$ violation in the Higgs potential}:  This case has
been discussed in \cite{pleitez2,cubic} in the context of neutrinoless double
beta decay and Majorana neutrino masses.  In general, when $L$ is violated
explicitly, it induces a non-zero triplet VEV $\langle T\rangle$ unless some
fine tuning is imposed.
\end{itemize}

\section{Reduction to the Standard Model}
The Higgs VEVs are arranged to first break $SU(3)_L\times U(1)_X$ to the SM
and then to break the SM.  This symmetry breaking hierarchy may be
represented as
\begin{equation}
SU(3)_L\times U(1)_X \stackrel{\langle\Phi\rangle}{\longrightarrow}
SU(2)_L\times U(1)_Y \stackrel{\langle\phi\rangle,\langle\phi'\rangle,
\langle H\rangle}{\longrightarrow} U(1)_Q \ .
\end{equation}
In this section, we consider the first stage of symmetry breaking and
examine the reduction of the 331 model to $SU(2)_L\times U(1)_Y$.

\subsection{331 symmetry breaking and gauge matching conditions}
When 331 is broken to the SM, the neutral gauge bosons $W_\mu^8$ and
$X_\mu$ mix to give the $Z_\mu'$ and hypercharge $B_\mu$ bosons.  In analogy
with the SM, we find
\begin{equation}
\pmatrix{B_\mu\cr Z'_\mu}
=
\pmatrix{\cos\theta_{331}&\sin\theta_{331}\cr
-\sin\theta_{331}&\cos\theta_{331}}
\pmatrix{W_\mu^8\cr X_\mu}\ ,
\end{equation}
where $\tan\theta_{331}=\sqrt{2}g/g_X$.  The hypercharge coupling constant
$g'$ is given from the gauge matching conditions (\ref{eq:match}) by
\begin{equation}
g'={1\over\sqrt{3}}g\cos\theta_{331}={1\over\sqrt{6}}g_X\sin\theta_{331}\ .
\end{equation}
Since $SU(3)_L\times U(1)_X$ is semi-simple, with two coupling constants, $g$
and $g_X$, the Weinberg angle is not fixed as it would be for unification
into a simple group.  However, the unknown coupling $g_X$ or equivalently
$\theta_{331}$ may be determined in terms of $\theta_W$.  We find
$\cos\theta_{331}=\sqrt{3}\tan\theta_W$, which gives
\begin{equation}
\alpha_X\equiv {g_X^2\over4\pi}=\alpha {6\over1-4\sin^2\theta_W}\ .
\end{equation}
This shows the interesting property that $\sin^2\theta_W<1/4$ with
$\sin^2\theta_W\approx1/4$ corresponding to strong coupling for the
$U(1)_X$ \cite{frampton,daniel}.  Although this is a tree level result, it
remains valid when the running of the coupling constants is taken into
account.  Since $\sin^2\theta_W(M_Z)=.233$ is already close to $1/4$ and
runs towards larger values as the scale is increased, this restriction gives
an absolute upper limit on the 331 breaking scale, $\mu\alt3$TeV.

Since this upper limit corresponds to infinite $\alpha_X$, more realistic
limits may be set by requiring the validity of perturbation theory.  Note,
however, that even at the $Z$-pole, we find a large $\alpha_X\approx0.7$
corresponding to $\sin^2\theta_{331}\approx0.09$.
Since $\alpha_X$ is large, it quickly runs to a Landau pole at around 3TeV
regardless of the 331
scale and indicates that a more complete theory may be necessary where the
$U(1)_X$ is embedded in a non-abelian group.

At this first stage of symmetry breaking, both dileptons and the $Z'$ gain
masses.  Assuming the $SU(2)_L$ subgroup remains unbroken, both members of
the dilepton doublet $(Y^{++},Y^+)$ gain identical masses.  Generalizing to
arbitrary Higgs representations for the moment, we find
\begin{eqnarray}
M_Y^2&=&{g^2\over2}\sum_i(C_2(R_i)-X_i^2/3)|\langle\chi_i\rangle|^2c_i
\nonumber\\
M_{Z'}^2&=&{2g^2\over3\sin^2\theta_{331}}\sum_i X_i^2
|\langle\chi_i\rangle|^2\ ,
\label{eq:masses}
\end{eqnarray}
where $R_i$ and $X_i$ denote the $SU(3)_L$ representation and $U(1)_X$
charge of the Higgs $\chi_i$.  $c_i=1$ for complex representations and $1/2$
for real ($X_i=0$) ones.
$C_2(R)$ is the quadratic Casimir of $SU(3)$ in representation
$R$, $T^aT^a=C_2(R) I$.

{}From (\ref{eq:masses}), we may define a generalization of the $\rho$
parameter,
\begin{equation}
\rho_{331}\equiv {M_Y^2\over M_{Z'}^2\sin^2\theta_{331}}
={3\over4}{ \sum_i(C_2(R_i)-X_i^2/3)|\langle\chi_i\rangle|^2c_i
\over \sum_i X_i^2 |\langle\chi_i\rangle|^2}\ .
\end{equation}
If there are more than one 331 breaking Higgs present, then their $X$
charges must be
chosen so as to preserve a common unbroken $SU(2)_L$ subgroup.  For an $SU(3)$
representation labeled by $(p,q)$, this may be done by picking $X=p-q$.
Using $C_2(p,q)={1\over3} (p^2+q^2+pq)+(p+q)$ in the standard normalization,
we find
\begin{equation}
\rho_{331}={3\over4}{\sum_{(p,q)}(p+q+pq)|\langle\chi_{(p,q)}\rangle|^2c_i
\over \sum_{(p,q)}(p-q)^2|\langle\chi_{(p,q)}\rangle|^2}\ .
\end{equation}

In the minimal 331 model, this symmetry breaking is accomplished by the
triplet Higgs $\Phi$ with $X=1$ ({\it i.e.}~$(p,q)=(1,0)$).  Defining
the 331 breaking VEV by
$\langle\Phi\rangle=u/\sqrt{2}$, we find $M_Y={g\over2}u$ and
$\rho_{331}=3/4$.  Since $\sin^2\theta_{331}\alt0.09$, the definition of
$\rho_{331}$ tells us that the $Z'$ must be considerably heavier
than the dileptons, $M_{Z'} \agt 3.9 M_Y$.  Demanding that
$\alpha_X(M_{Z'})<2\pi$ gives the upper limit $M_{Z'}<2.2$TeV, and hence
$M_Y<430 (\sqrt{4\rho_{331}/3})$GeV for the masses of the new gauge
bosons\cite{phenom}.

Lower bounds on the dilepton mass have be studied in
\cite{su15fn,fujii,fujii2,carlson}.  The best current lower bound
comes from polarized muon decay \cite{carlson} which is especially sensitive
to a non-standard charged-current interaction \cite{fn:bound}.
At 90\% C.L., we find
$M_Y>300$GeV \cite{phenom} with a corresponding limit $M_{Z'}>1.4
(\sqrt{3/4\rho_{331}})$TeV on the $Z'$ mass.

The imposition of both lower and upper limits on the scale of 331 physics is
very constraining.
Although larger values of $\rho_{331}$ coming from a non-minimal
Higgs sector would relax these bounds \cite{phenom}, the range of new
physics is still
limited to within about one order of magnitude above the $Z$-pole.
As a result this model has the positive feature that it is easily testable.

\subsection{reduction of the Higgs sector}
We now focus on the minimal Higgs sector, given by the three $SU(3)_L$
triplets, (\ref{eq:tripletH}), and the $SU(3)_L$ sextet, (\ref{eq:sextetH}).
The most general scalar potential involving these fields is given by
\begin{equation}
V(\Phi,\phi,\phi',H)=V^{(2)}+V^{(3)}+V^{(4a)}+\cdots+V^{(4e)}\ ,
\end{equation}
where
\begin{eqnarray}
V^{(2)} =&& \mu_1^2\Phi^\dagger\Phi+\mu_2^2\phi^\dagger\phi
     +\mu_3^2\phi'^\dagger\phi'+\mu_4^2{\rm Tr}\,H^\dagger H\nonumber\\
V^{(3)} =&& \alpha_1\Phi\phi\phi'+\alpha_2(\Phi^T H^\dagger\phi')
   +\alpha_3 (\phi^T H^\dagger\phi)+\alpha_4HHH+{\rm h.c.}\nonumber\\
V^{(4a)} =&& a_1(\Phi^\dagger\Phi)^2+a_2(\phi^\dagger\phi)^2+
    a_3(\phi'^\dagger\phi')^2+a_4(\Phi^\dagger\Phi)(\phi^\dagger\phi)
   +a_5(\Phi^\dagger\Phi)(\phi'^\dagger\phi')
   +a_6(\phi^\dagger\phi)(\phi'^\dagger\phi')\nonumber\\
 &&+a_7(\Phi^\dagger\phi)(\phi^\dagger\Phi)
   +a_8(\Phi^\dagger\phi')(\phi'^\dagger\Phi)
   +a_9(\phi^\dagger\phi')(\phi'^\dagger\phi)
   +[a_{10}(\Phi^\dagger\phi)(\phi'^\dagger\phi)+{\rm h.c.}]\nonumber\\
V^{(4b)} =&& b_1\Phi^\dagger H\Phi\phi+b_2\phi'^\dagger H\phi'\phi
  +b_3\phi^\dagger H \Phi\phi'+{\rm h.c.}\nonumber\\
V^{(4c)} =&& c_1\phi\phi H H+c_2\Phi\phi' H H + {\rm h.c.}\nonumber\\
V^{(4d)} =&& d_1(\Phi^\dagger\Phi){\rm Tr}\,H^\dagger H
  +d_2(\Phi^\dagger HH^\dagger\Phi)
  +d_3(\phi^\dagger\phi){\rm Tr}\,H^\dagger H
  +d_4(\phi^\dagger HH^\dagger\phi)\nonumber \\
&&+d_5(\phi'^\dagger\phi'){\rm Tr}\,H^\dagger H
  +d_6(\phi'^\dagger HH^\dagger\phi')\nonumber\\
V^{(4e)}=&&e_1({\rm Tr}\,H^\dagger H)^2
  + e_2 {\rm Tr}\,H^\dagger H H^\dagger H \ .
\end{eqnarray}
The quartic terms, $V^{(4a)},\ldots,V^{(4e)}$, have been broken up according
to the $SU(3)$ representation contents,
$({\bf3}\times {\bf3}\times {\bf3}^* \times {\bf3}^*)$,
$({\bf3}\times {\bf3}\times {\bf3}^* \times {\bf6})$,
$({\bf3}\times {\bf3}\times {\bf6}\times {\bf6})$,
$({\bf3}\times {\bf3}^*\times {\bf6}\times {\bf6}^*)$ and
$({\bf6}\times {\bf6}\times {\bf6}^*\times {\bf6}^*)$
respectively.

According to the previously worked out lepton number assignment, the
terms $\alpha_3$, $\alpha_4$, $a_{10}$, $b_3$ and $c_2$ violate lepton
number explicitly.  Soft lepton number violation may be accomplished by
setting $\alpha_3$,~$\alpha_4\ne0$ \cite{pleitez2,cubic}.  Since we are
presently interested in the minimal 331 model where lepton number is
not violated, we instead take $\alpha_3=\alpha_4=a_{10}=b_3=c_2=0$.  In
addition, the remaining parameters must be chosen so that the $SU(2)_L$
triplet $T$ does not develop a VEV and hence break lepton number
spontaneously.  As we have discussed in the previous section, this
theory is not a complete theory.  Thus lepton number conservation may
be a consequence of physics beyond the 331 model.

The first stage of symmetry breaking is governed by the triplet $\Phi$ with
potential
\begin{eqnarray}
V&=&\mu_1^2\Phi^\dagger\Phi+a_1(\Phi^\dagger\Phi)^2+\cdots\nonumber\\
&=&a_1(\Phi^\dagger\Phi-u^2/2)^2+\cdots\ ,
\end{eqnarray}
where $\langle\Phi\rangle=u/\sqrt{2}=\sqrt{-\mu_1^2/2a_1}$ (with $u$ chosen
to be real).  Of the original
six real degrees of freedom, five become the longitudinal modes of the
dileptons and the $Z'$, leaving the physical heavy $SU(2)_L$ singlet
$\sqrt{2}{\rm Re}\,\varphi^0$ with mass $M^2=-2\mu_1^2=2a_1u^2$.  The
singlets $\Delta^-$ and $\rho^{--}$ also become heavy with masses
$M_{\Delta^-}^2=a_7u^2/2$ and $M_{\rho^{--}}^2=a_8u^2/2$.

The decomposition of the sextet $H$ is a bit trickier.  Due to the term $d_2$,
we expect the masses to obey $M_T^2<M_{\Phi_3}^2<M_{\eta^{--}}^2$, equally
spaced with $\Delta M^2=d_2u^2/4$.  In this case, the $SU(2)_L$ triplet is
naturally light, with $\Phi_3$ and $\eta^{--}$ heavy.  However, this is
unappealing since $H$ was introduced in the first place so the charged
leptons may get their masses from $\langle\Phi_3\rangle$.  Thus we need to
set $d_2\approx0$, with the consequence that both $T$ and $\eta^{--}$ may be
light \cite{fn:Tvev}.

After 331 breaking, the resulting scalars take the form of a three Higgs
doublet model with the additional light fields $T$ and $\eta^{--}$.  For the
three Higgs doublets only, we find the tree level reduced potential
\begin{eqnarray}
V_{3\scriptscriptstyle \rm HD}(\Phi_i)=
&&\phantom{+}\sum_i m_i^2(\Phi_i^\dagger\Phi_i^{\vphantom{\dagger}})
+\sum_{i<j}[ m_{ij}^2(\Phi_i^\dagger\Phi_j^{\vphantom{\dagger}})
+{\rm h.c.}]\nonumber\\
&&+\sum_{i\le j}\lambda_{ij}(\Phi_i^\dagger\Phi_i^{\vphantom{\dagger}})
	(\Phi_j^\dagger\Phi_j^{\vphantom{\dagger}})
+\sum_{i<j}\lambda_{ij}'(\Phi_i^\dagger\Phi_j^{\vphantom{\dagger}})
	(\Phi_j^\dagger\Phi_i^{\vphantom{\dagger}})\nonumber\\
&&+[\lambda_{1313}(\Phi_1^\dagger\Phi_3^{\vphantom{\dagger}})
	(\Phi_1^\dagger\Phi_3^{\vphantom{\dagger}})
+\lambda_{1223}(\Phi_1^\dagger\Phi_2^{\vphantom{\dagger}})
	(\Phi_2^\dagger\Phi_3^{\vphantom{\dagger}})+{\rm h.c.}]\ .
\end{eqnarray}
A completely general three Higgs doublet potential includes additional
possible terms in the last line.  However, since the model was originally
$SU(3)_L\times U(1)_X$ invariant, only the ones explicitly shown
here are present at tree level.  The coefficients are given by
\begin{equation}
\begin{array}[b]{rclrcl}
m_1^2&=&\mu_2^2+a_4u^2/2\qquad&\qquad m_{12}^2&=&-\alpha_1^*u/\sqrt{2}\\
m_2^2&=&\mu_3^2+a_5u^2/2&m_{13}^2&=&-b_1^*u^2/2\sqrt{2}\\
m_3^2&=&\mu_4^2+d_1u^2/2&m_{23}^2&=&\alpha_2u/2\\[4pt]
\lambda_{11}&=&a_2-a_4^2/2a_1&\lambda_{1313}&=&-c_1^*\\
\lambda_{22}&=&a_3-a_5^2/2a_1&\lambda_{1223}&=&b_2^*/\sqrt{2}\\
\lambda_{33}&=&e_1+e_2/2-d_1^2/2a_1\\[4pt]
\lambda_{12}&=&a_6+a_9-a_4a_5/2a_1&\lambda_{12}'&=&-a_9\\
\lambda_{13}&=&d_3+d_4/2-a_4d_1/2a_1&\lambda_{13}'&=&-d_4/2\\
\lambda_{23}&=&d_5-a_5d_1/2a_1&\lambda_{23}'&=&d_6/2\ .
\end{array}
\end{equation}
In performing the dimensional reduction,
we have assumed $\alpha_1,\alpha_2\sim v^2/u$ and $b_1\sim v^2/u^2$
are small where $v$ is an $SU(2)_L$ breaking VEV.  This assumption is
necessary to ensure $m_{ij}^2\sim v^2$ and hence to preserve the symmetry
breaking hierarchy.

Three Higgs doublet models have been studied previously, usually in the
context of the Weinberg model of CP violation
\cite{weinberg,deshpande,branco}.
However, in this case $V_{3\scriptscriptstyle\rm HD}$ is {\it not} invariant
under $\Phi_i\to-\Phi_i$ which is often imposed to enforce natural flavor
conservation (NFC)\cite{glashow}.  Although it is possible to eliminate the
$m_{ij}^2$ terms by a unitary rotation of the $\Phi_i$'s, doing so would
complicate the equations by introducing additional quartic
couplings and would also affect the Yukawa couplings.  Thus we find it
more convenient to leave these off-diagonal terms in
$V_{3\scriptscriptstyle\rm HD}$.

In the absence of NFC there may be large FCNH
processes.  Since the $\Phi_i$ are remnants of the original
$SU(3)_L\times U(1)_X$ invariant fields, their couplings are restricted over
that of a generic $SU(2)_L\times U(1)_Y$ three Higgs doublet model.
However, we find that these additional constraints are insufficient to
implement NFC.  In the quark sector, this should come as no surprise because
the third family is explicitly different, resulting in both $Z'$ mediated
FCNC in the gauge sector and FCNH in the scalar sector.  In the leptonic
sector, both $\Phi_1$ and $\Phi_3$ may couple to leptons, resulting in FCNH
and lepton flavor violation.  However, since the leptons are treated
identically, it is possible to impose an additional discrete symmetry
that allows only a single Higgs to couple to the leptons.  This possibility
is explored further in the next section.

Because $T$ and $\eta^{--}$ carry lepton number, they do not mix with the
three doublets (in the absence of lepton number violation).  Analysis of the
scalar potential indicates that a stable minimum with $\langle T\rangle=0$
can be found for large regions of parameter space \cite{foot,pleitez2}.
As long as $T$ does not pick up a VEV, both $T$ and $\eta^{--}$ have no
effect on symmetry breaking of the SM.  This allows us to ignore these
additional scalars and only focus on the three Higgs doublets of the 331
model.

\subsection{Higgs sector CP violation}
There are several options for CP violation in the 331 model.  With complex
Yukawa couplings, hard CP violation occurs through the CKM phase.  In
addition to the ordinary CKM coupling of the $W$ charged current to quarks,
the 331 model also has dilepton charged current couplings.  This leads to
new mixing angles as well as additional CP violating phases in both the
leptonic and hadronic sector.  This is perhaps the most straightforward
generalization of CP violation in the SM.  However, the additional phases
may lead to novel effects such as large lepton EDMs
which are otherwise undetectably small in the SM.

CP violation may also occur in the extended Higgs sector \cite{lee,weinberg}.
For three Higgs doublets, CP violation may be either explicit (complex
$m_{ij}^2$, $\lambda_{1313}$ and $\lambda_{1223}$ in
$V_{3\scriptscriptstyle\rm HD}$) or spontaneous.  In both cases, CP violation
occurs through charged and neutral Higgs exchange.  The original motivation
for introducing three doublets to the SM was to obtain CP violation in the
scalar sector without FCNH.  On the other hand, the 331 model {\it has} FCNH
but requires three doublets for mass generation.  In this case,
CP violation from tree level FCNH cannot be ignored \cite{lee,liuw}.
In addition, since the new triplet and singlet $T$ and $\eta^{--}$ couple to
leptons, they may also contribute to leptonic CP violation as discussed
in Ref.~\cite{zee}.

\subsection{Standard Model breaking}
When $m_i^2$, $m_{ij}^2<0$ in $V_{3\scriptscriptstyle\rm HD}$, the three
Higgs doublets pick up (possibly complex) VEVs $\langle
\Phi_i\rangle=v_i/\sqrt{2}$ and breaks $SU(2)_L\times U(1)_Y$.  The
resulting physical scalars are four charged Higgs, $H_{1,2}^\pm$, and five
neutral ones $h_{1,\ldots,5}^0$.  The physical states
$H_{1,2}^+$ and the Goldstone mode are related to the original $\phi^+_i$
via a $3\times3$ unitary matrix with a single physical CP violating angle
(distinct from the usual CKM angle) \cite{albright}.  CP violation
in the neutral Higgs sector manifests itself in the mixing of the CP even
and CP odd scalars.

While the other light scalars $T$ and $\eta^{--}$ have no effect on symmetry
breaking, they acquire masses related to the VEVs $v_i$.  Because $SU(2)_L$
is broken, the triplet will become split in mass and $T^{++}$ and $\eta^{++}$
will mix.
This second stage of symmetry
breaking will also have an effect on the $SU(3)_L$ particles.  In
particular, the dilepton doublet will become split in mass and the
$Z$ and $Z'$ will mix.  Expressions for all tree level gauge boson masses
and $Z$--$Z'$ mixing have been given in \cite{daniel}.  Because of the
symmetry breaking hierarchy, these effects may be considered as
perturbations to the results where $SU(2)_L$ remains unbroken.  However, in
the 331 model, this must often be treated with care since the two scales are
within an order of magnitude of each other.

\section{Lepton flavor violation and CP violation}
We now turn to the leptonic sector of the 331 model.  Since the leptons are
in the ${\bf3}^*_0$ representation of $SU(3)_L\times U(1)_X$, the lepton
bi-linear $\psi\psi$ transforms as ${\bf3}^*_0\times{\bf3}^*_0=
{\bf3}_0+{\bf6}^*_0$.  Thus leptons may have gauge invariant Yukawa
couplings to the triplet $\phi$ and sextet $H$.  We write the Yukawa
interaction as
\begin{equation}
-{\cal L}={1\over\sqrt{2}}
\overline{\psi_i'^\alpha}h_s^{ij} \psi_j'^{\beta c} H^*_{\alpha\beta}
- {1\over2} \overline{\psi_i'^\alpha}h_a^{ij}
\psi_j'^{\beta c} \phi^\gamma\epsilon_{\alpha\beta\gamma}+{\rm h.c.}\ ,
\label{eq:lepyuk}
\end{equation}
where the primes denote weak eigenstates.  Here, $i,j$ are family indices
and $\alpha,\beta,\gamma=1,2,3$ are $SU(3)$ group indices.  From the symmetry
properties of (\ref{eq:lepyuk}), the Yukawa coupling matrix $h_s$ is
symmetric and $h_a$ is antisymmetric.  The above factors have been chosen
so the charged lepton mass matrix will take on a simple form and differs
from the convention used in \cite{cubic}.

We may rewrite Eq.~(\ref{eq:lepyuk}) in terms of $SU(2)_L$ component fields.
Using the definitions of (\ref{eq:tripletH}) and (\ref{eq:sextetH}), the
Yukawa interactions may be written
\begin{eqnarray}
-{\cal L}&=&\overline{L_L'}[h_s\Phi_3+h_a\Phi_1]e_R'\nonumber\\
&&\ +{1\over\sqrt{2}}\overline{L_L'}h_s\widetilde{T}L_L'^c
-{1\over2}\overline{L_L'}h_a(i\tau^2)L_L'^c\Delta^-
+{1\over\sqrt{2}}\overline{e_R'^c}h_se_R'\eta^{++}+{\rm h.c.}\ ,
\end{eqnarray}
where the family indices have been suppressed and $L_L=(\nu,\ell^-)_L$ is the
SM lepton doublet.  The first line gives a two Higgs doublet SM interaction
and the second line gives the interaction with new 331 scalars.  While
$\Delta^-$ is heavy, $T$ and $\eta^{++}$ may be light, and resemble the
scalars introduced in Ref.~\cite{zee} for generating leptonic CP violation
\cite{fn:notzee}.
As we noted before, this model does not satisfy the requirements for NFC and
hence violates lepton family number via FCNH.  However, unlike a general
two Higgs doublet model with arbitrary Yukawa couplings,
$SU(3)_L$ gauge invariance restricts the form of $h_s$ and $h_a$.
This has important consequences as shown below.

\subsection{lepton masses and mixing}
When the SM is broken by the Higgs doublet VEVs
$\langle\Phi_i\rangle=v_i/\sqrt{2}$, the charged leptons get a mass matrix
$M_\ell=(h_sv_3+h_av_1)/\sqrt{2}$.  Since $h_s$ ($h_a$) is (anti-)symmetric,
$M_\ell$ is an arbitrary complex $3\times3$ matrix.  We diagonalize this
matrix by a bi-unitary transformation $E_L^\dagger M_\ell
E_R^{\vphantom{\dagger}}={\rm diag}(m_e,m_\mu,m_\tau)$.  As a result,
physical (mass) eigenstates
are related to the weak eigenstates according to
\begin{equation}
e_L'=E_Le_L\qquad e_R'=E_Re_R\qquad\nu_L'=F_L\nu_L\ ,
\end{equation}
where we also introduce a unitary transformation for the neutrinos.

In terms of the physical basis, the $W$ and dilepton charged currents become
\begin{equation}
\begin{array}{lclcl}
J_+^\mu&=&\phantom{-}\overline{\nu}\gamma^\mu\gamma_L
[F_L^\dagger E_L^{\vphantom{\dagger}}]e
&=&\phantom{-}\overline{\nu}\gamma^\mu\gamma_LV_We\\
J_{Y^+}^\mu&=&\phantom{-}\overline{e^c}\gamma^\mu\gamma_L
[E_R^TF_L^{\vphantom{T}}]\nu
&=&\phantom{-}\overline{e^c}\gamma^\mu\gamma_L
V_Y^{\vphantom{\dagger}}V_W^\dagger\nu\\
J_{Y^{++}}^\mu&=&-\overline{e^c}\gamma^\mu\gamma_L
[E_R^TE_L^{\vphantom{T}}]e
&=&-\overline{e^c}\gamma^\mu\gamma_LV_Ye\ ,
\label{eq:lepcc}
\end{array}
\end{equation}
where $V_W=F_L^\dagger E_L^{\vphantom{\dagger}}$ and $V_Y=E_R^T
E_L^{\vphantom{T}}$ are unitary mixing matrices in
the leptonic sector.  Thus we find that in addition to a possible
leptonic CKM mixing coming from massive neutrinos, lepton family number
may also be violated in the interaction with dileptons.
Note that the current $J_{Y^{++}}$ in (\ref{eq:lepcc}) may be rewritten as
$J_{Y^{++}}^\mu=-{1\over2} \overline{e^c}\gamma^\mu
(V_Y^{\vphantom{T}}\gamma_L-V_Y^T\gamma_R)e$, showing that the doubly
charged dilepton has both left- and right-handed couplings and that
the family diagonal coupling is purely axialvector.

If the neutrinos are massless, then we may pick $F_L=E_L$, or equivalently
$V_W=1$.  In this case, the ordinary $W$ charged current is family diagonal,
and the dilepton interaction is determined completely by $V_Y$.  In general,
a $3\times3$ unitary matrix is fixed by three angles and six phases.
Unlike the
normal CKM case, because $V_Y$ is determined entirely from the charged lepton
sector, we may only rotate away 3 phases, corresponding to $E_{L,R}\to
E_{L,R}K$ (where $K$ is a diagonal matrix of phases) which preserves the
reality of the diagonal charged lepton masses.  As a result, $V_Y$ depends
on a total of six real parameters: three angles and three phases.

If the triplet $T$ gets a VEV, then the neutrinos pick up a Majorana mass
$M_\nu=\sqrt{2}h_s\langle T\rangle$.  Neutrino masses may also arise by
adding right-handed neutrino states.  In both cases, $F_L$ must then be
chosen to diagonalize the neutrino mass matrix.  For Majorana neutrinos,
$M_\nu$ is symmetric and we can find $F_L$ such that $F_L^\dagger M_\nu
F_L^*$ is diagonal.  In general, diagonalization may be more complicated.

With massive neutrinos, $V_W$ describes mixing in the ordinary leptonic
sector.  The number of possible CP violating phases depends on the nature of
the neutrinos.  For Majorana neutrinos, if $V_Y$ is
fixed as above to have three angles and three phases, then there is no
more freedom to rotate away any phase because of the Majorana nature of
the neutrinos.  Hence, there are three angles and six phases in $V_W$.
On the other hand, we may choose to rotate away
three phases in $V_W$ by redefining the charged lepton phases,
leaving $V_W$ with three angles and three phases
and $V_Y$ with three angles and six phases.  In both cases, there are a
total of nine possible CP violating phases.  Physically, there should be no
difference between these cases, so we may choose to distribute the phases
among the various charged currents in the most convenient manner.
For Dirac neutrinos, we may remove three additional phases, leaving a total
of six CP violating phases.  A possible distribution of phases is one in
$V_W$ and five in $V_Y$, so that $V_W$ has the usual form for the Dirac
case.

While nine, or even six, CP violation phases may seem like a lot, in many
specific 331 models of neutrino mass, the neutrino mass matrices are
related to the charged lepton mass matrices, and hence lead to relations
among the mixing angles and phases.
Thus the number of independent phases may be no larger than three, the
minimum coming from the doubly charged dilepton current.  In particular,
for Majorana neutrinos that get masses from the {\it same} Yukawa couplings
$h_s$ and $h_a$, there is no additional freedom, and the matrix $V_W$ may
be specified in terms of the six parameters of $V_Y$, although the exact
relation is usually rather complicated \cite{cubic}.

\subsection{dilepton mediated rare lepton decays}
Even with massless neutrinos, the doubly charged dilepton may have family
non-diagonal interactions because of the new mixing given by $V_Y$.
As a result, lepton flavor violating processes such as $\mu\to3e$ and
$\mu\to e\gamma$ may occur.  In addition, the phases in $V_Y$ lead to
leptonic CP violation which may be observed by detecting a triple product
correlation in $\mu\to3e$ \cite{treiman} decay or by measuring non-zero
lepton EDMs.  Since these exotic decays have not been seen, this leads to
strong constraints on the allowed mixing coming from $V_Y$.

The decay $\mu\to3e$ proceeds via tree level dilepton exchange as shown in
Fig.~\ref{fig:mu3e}.  Ignoring
final state particle masses, we find
\begin{equation}
{{\rm BR}(\mu\to3e)\over{\rm BR}(\mu\to e\overline{\nu_e}\nu_\mu)}
=\left(M_W\over M_Y\right)^4|V_Y^{11}|^2
(|V_Y^{12}|^2+|V_Y^{21}|^2)\ ,
\label{eq:brmu3e}
\end{equation}
and similar expressions for the processes $\tau\to3\mu$,
$\tau^-\to\mu^+e^-e^-$, and
$\tau^-\to e^+\mu^-\mu^-$ with the appropriate replacement of the family
indices.  For $\tau^-\to e^-\mu^+\mu^-$ and $\tau^-\to\mu^-e^+e^-$, the family
diagonal coupling $|V_Y^{11}|^2$ must be replaced by the appropriate
off-diagonal coupling $|V_Y^{i3}|^2+|V_Y^{3i}|^2$ with $i=2,1$ respectively.
The present experimental limits are \cite{pdb}
\begin{eqnarray}
{\rm BR}(\mu\to3e)&<&1.0\times10^{-12}\nonumber\\
{\rm BR}(\tau\to3\ell)&<&3.4\times10^{-5}\ ,
\label{eq:mu3e}
\end{eqnarray}
(at 90\% C.L.), where $\ell$ denotes either $\mu$ or $e$.  The constraints
for the various $\tau\to3\ell$ channels are given in \cite{pdb} and are all
less than the order of $10^{-5}$.  Clearly the experimental bounds are not
as well determined for $\tau$ decay as it is for $\mu$ decay.  This allows
for relatively large $e$--$\tau$ and $\mu$--$\tau$ mixing, with important
consequences for the electron and muon EDM.

A standard method for suppressing flavor changing processes is to make the
exchanged particle very heavy.  However, in the present case there is an
upper limit on the dilepton mass, $M_Y<430$GeV (in the minimal case where
$\rho_{331}=3/4$).  As a result, we can restrict the mixing allowed by
$V_Y$.  Assuming the lepton families are almost diagonal, $V_Y\approx1$,
we may write $V_Y^{ij}=\delta^{ij}+2\alpha^{ij}e^{i\theta_{ij}}$ in the
small mixing approximation where $\alpha^{ij}=-\alpha^{ji}$ are the three
mixing angles and $\theta_{ij}=-\theta_{ji}$ the three CP violating
phases of $V_Y$.  In the Appendix, we show how $\alpha$ and $\theta$ may be
related to the original Yukawa couplings $h_s$ and $h_a$ of
(\ref{eq:lepyuk}).  In terms of this parametrization, the
experimental bounds (\ref{eq:mu3e}) give the limits
\begin{eqnarray}
|\alpha^{12}|&<&1.0\times10^{-5}\nonumber\\
|\alpha^{13}|&<&0.096\nonumber\\
|\alpha^{23}|&<&0.096\ ,
\label{eq:choice1}
\end{eqnarray}
justifying the small mixing approximation, at least for the first two
families.

Curiously, there is a second choice for $V_Y$ consistent with the above
limits.  In this case, $Y^{--}$ has a mostly off-diagonal coupling to the
first two families, $Y^{--}\to e^-\mu^-$, or, in terms of the mixing
matrix, $|V_Y^{12}|\approx|V_Y^{21}|\approx1$.  The other components are
restricted by
\begin{eqnarray}
|V_Y^{11}|^2&<&4.1\times10^{-10}\nonumber\\
|V_Y^{13}|^2+|V_Y^{31}|^2&<&0.062\nonumber\\
|V_Y^{23}|^2+|V_Y^{32}|^2&<&0.062\ ,
\label{eq:choice2}
\end{eqnarray}
and $|V_Y^{22}|^2\alt10^{-3}$ from unitarity of $V_Y$.  This large
mixing case corresponds to $\alpha^{12}\approx\pi/4$, and occurs in the
limit when the diagonal Yukawa couplings are identical, $h_s^1=h_s^2$,
with the result that $m_e, m_\mu =(h_s^1|v_3|\pm y^{12}|v_1|)/\sqrt{2}$.
The third family has the standard diagonal form, $m_\tau=h_s^3|v_3|/\sqrt{2}$.

It is easy to show that these two cases are the only possible solutions
consistent with (\ref{eq:mu3e}).  Furthermore, these limits on $V_Y$ are
independent of any neutrino masses and mixing.  However, the second case may
be marginally ruled out from an analysis of transverse electron polarization
in muon decay, as we indicate below.  On the theoretical side, as well, there
appears to be no
principle which would enforce the necessary equality between $h_s^1$
and $h_s^2$.  Thus the second case will not be further investigated.

Lepton flavor violating processes of the form
$\mu\to e\gamma$ may also occur via either $W^-$, $Y^-$ or $Y^{--}$ exchange
at one-loop.  For both singly charged cases, a neutrino is running in the
loop, and hence the amplitude vanishes for massless neutrinos.  For massive
neutrinos, the GIM cancellation is not perfect, but nevertheless leads to a
large suppression of the amplitude.  On the other hand, since the $Y^{--}$
has both right- and left-handed couplings, it leads to a large contribution
to $\mu\to e\gamma$ as shown in Fig.~\ref{fig:muegamma}.

Assuming the intermediate charged leptons are light, $m_i\ll M_Y$, the one
loop diagrams lead to transition magnetic and electric dipole moments
\begin{equation}
\mu_{12},d_{12}={3e G_F\over4\sqrt{2}\pi^2}\left(M_W\over M_Y\right)^2
\sum_i(V_Y^{1i}V_Y^{i2\,*}\pm V_Y^{i1}V_Y^{2i\,*})m_i\ ,
\end{equation}
resulting in a decay width of
\begin{equation}
\Gamma_{\mu\to e\gamma}={m_\mu^3\over8\pi}(|\mu_{12}|^2+|d_{12}|^2)\ ,
\end{equation}
(ignoring the electron mass).  Since $\alpha^{12}\ll1$, the intermediate
state $\tau$ dominates, leading to a branching ratio
\begin{equation}
{\rm BR}(\mu\to e\gamma)=
{54\alpha\over\pi}\left(M_W\over M_Y\right)^4 \left(m_\tau\over m_\mu\right)^2
(|V_Y^{13}|^2|V_Y^{32}|^2+|V_Y^{31}|^2|V_Y^{23}|^2)\ .
\end{equation}
Compared to $\mu\to3e$ decay, Eq.~(\ref{eq:brmu3e}), the loop factor
$\alpha/\pi$ is compensated for by the larger phase space and the heavy
$\tau$.  Using the upper limit on $M_Y$ and the experimental limit ${\rm
BR}(\mu\to e\gamma)_{\rm expt}<4.9\times10^{-11}$ \cite{pdb}, we find
\begin{equation}
|\alpha^{13}\alpha^{23}|<5.9\times10^{-6}\ ,
\end{equation}
a combined limit much stronger than the individual ones of
Eq.~(\ref{eq:choice1}).

\subsection{lepton electric dipole moments}
In addition to large transition dipole moments, one-loop diagrams similar to
those of Fig.~\ref{fig:muegamma} may lead to large EDMs.  The electron
EDM is calculated to be
\begin{eqnarray}
d_e&=&{3e G_F\over2\sqrt{2}\pi^2}\left(M_W\over M_Y\right)^2
\sum_i{\rm Im}\,(V_Y^{1i}V_Y^{i1\,*})m_i\\
&\approx&-{3\sqrt{2}e G_F\over\pi^2}\left(M_W\over M_Y\right)^2
\sum_i m_i|\alpha^{1i}|^2\sin2\theta_{1i}\ ,
\end{eqnarray}
and similarly for $d_\mu$ and $d_\tau$.  We observe that $Y^{--}$ mediated
CP violation occurs {\it only} through lepton flavor changing interactions.
Putting in numbers, we estimate
\begin{eqnarray}
d_e&\approx&8.5|\alpha^{13}|^2\sin2\theta_{13} \times10^{-21}e\;{\rm cm}\\
d_\mu&\approx&8.5|\alpha^{23}|^2\sin2\theta_{23} \times10^{-21}e\;{\rm cm}\ ,
\label{eq:edm}
\end{eqnarray}
where terms proportional to $|\alpha^{12}|^2$ ($<10^{-10}$ from
Eq.~(\ref{eq:choice1})) have been ignored.  The estimate for $d_e$ is
extremely
large compared to the experimental limit $|d_e|<1.9\times10^{-26}e\;{\rm cm}$
\cite{abdullah} but depends on undetermined $e$--$\tau$ mixing parameters.

An interesting consequence of having only off-diagonal CP violating
interactions is the inverse relation $d_\mu/d_\tau\approx-m_\tau/m_\mu$.
While any observed EDM would indicate physics beyond the SM (which
predicts unobservably small lepton EDMs \cite{bernreuther}), this relation
may be of use in verifying the 331 model of CP violation.

In principle, CP violation may also show up in ordinary muon decay due to
interference between the $W^-$ and $Y^-$ induced amplitudes.  In the
presence of lepton flavor violation, the unobserved final state neutrinos
may be in any family.  Nevertheless, this is easily taken into account
\cite{langacker}, and does not affect the investigation of polarized muon
decay in Ref.~\cite{carlson}.  For non-diagonal $V_Y$, the
muon decay transverse polarization parameters $\beta$ and $\beta'$
\cite{kinoshita} become non-zero,
\begin{eqnarray}
\left.\beta\atop\beta'\right\} &=& -8 \left(M_W\over M_Y\right)^2
\left\{\rm Re\vphantom{\beta}\atop Im\right\} (V_Y^{12}V_Y^{21*})\nonumber\\
&\approx&32\left(M_W\over M_Y\right)^2 |\alpha^{12}|^2
\left\{\cos2\theta_{12}\atop\sin2\theta_{12}\right\}\ .
\end{eqnarray}
In practice, this indication of CP violation in muon decay is unobservable,
as it is proportional to the very small $\mu$--$e$ mixing.  We predict
$\beta'/A\alt10^{-11}$ where $A=16(1+(M_W/M_Y)^4)\approx16$ normalizes the
decay rate.  This is some eight orders of magnitude below current experimental
limits \cite{burkard}.
On the other hand, had there been large mixing, as in (\ref{eq:choice2}), we
would have found $|\beta/A|$,~$|\beta'/A| \sim {1\over2}(M_W/M_Y)^2 \geq
0.017$ which is ruled out by experiment at 90\% C.L.

So far we have only considered lepton flavor changing processes mediated by
dilepton gauge bosons.  In general, scalar exchange will also contribute to
both lepton flavor violation and CP violation.  However, since the lepton
Yukawa couplings are very small, these superweak interactions are often
negligible compared to the dilepton interaction.  Only in the absence of
lepton flavor violation will the scalar sector play an important role in
CP violation.

\subsection{elimination of lepton flavor violation}
In order to suppress lepton flavor violation, the dilepton mixing angles
$\alpha$ must be very small.  From the appendix, we see that this means
the anti-symmetric Yukawa coupling needs to be very small,
$h_a|v_1|\ll h_s|v_3|$.  We now have a naturalness problem since
the limits on $\mu$--$e$ transitions require $h_a$ to be about five orders
of magnitude less than
$h_s$ (which is already small to accommodate the observed lepton masses).
One solution to this problem is to simply set $h_a=0$ which can be enforced
by a discrete symmetry $\phi\to-\phi$ (along with an appropriate transformation
of the quark fields).
This discrete symmetry actually serves two purposes.  It prevents the
doubly charged dilepton from having family non-diagonal couplings and
prevents FCNH by allowing only a single Higgs multiplet (the sextet) to
couple to the leptons.  With massless neutrinos, this symmetry prevents
$\Delta L_i=\pm1$ lepton flavor violation (although $\Delta L_i=\pm2$ would
still be allowed).

Since dilepton mediated CP violation occurs through $\Delta L_i=\pm1$
interactions, it is also eliminated by this discrete symmetry, leaving
CP violation to the scalar sector.  With massless neutrinos in the three
Higgs doublet model, CP violation only occurs through mixing of the CP
even and odd neutral Higgs.  Because the Yukawa couplings are proportional
to the charged lepton masses, $h_s\sim m_\ell/M_W$, the one-loop
contribution to the lepton EDM is proportional to the cube of the lepton mass,
\begin{equation}
d_\ell\simeq{e\sqrt{2}G_Fm_\ell^3\over8\pi^2M^2}\ln\left(m_\ell\over
M\right)^2 \delta\ ,
\label{eq:sedm}
\end{equation}
where $M$ and $\delta$ are the effective scalar mass and mixing.

Another source of CP violation, briefly touched upon above, is the mixing of
the 331 scalars $T^{++}$ and $\eta^{++}$.  Since the unmixed scalars couple
to leptons of different chirality, large CP violating effects are
proportional to the amount of singlet-triplet mixing as well as their mass
splitting.  The one-loop EDM induced by $T^{++}$--$\eta^{++}$ mixing is
again proportional to $m_\ell^3$, giving the same estimate,
Eq.~(\ref{eq:sedm}), but this time reduced by a factor $\delta M^2/M^2$
where $\delta M^2$ is the singlet--triplet mass splitting.

While both scalar one-loop contributions to the electron EDM are proportional
to the electron mass cubed and hence very small, two-loop contributions have
been shown to be important \cite{barr} and can lead to a fairly large electron
EDM, albeit still smaller than the dilepton loop result (\ref{eq:edm}).  The
two-loop contribution also dominates for the muon EDM, but the $\tau$ is
sufficiently heavy that the one-loop contribution may be more important in that
case.  Assuming large CP violation in the scalar sector and a typical scalar
mass of 100GeV leads to the order of magnitude estimates $d_e\sim10^{-27}$,
$d_\mu\sim10^{-25}$ and $d_\tau\sim10^{-23}$ $e\;{\rm cm}$.  This
prediction is similar to that of other flavor conserving scalar models of CP
violation \cite{weinberg,barr,twoloop}.

\section{Conclusion}
We have seen that in the general 331 model the leptons gain mass via
symmetric and anti-symmetric couplings to two Higgs doublets.  This leads to
the possibility of both FCNH and lepton flavor violation mediated by
dilepton exchange.  In addition to neutrino mixing, there are nine
physical parameters in the leptonic sector:  three masses $m_i$,
three mixing angles $\alpha^{ij}$ and three CP violating phases
$\theta_{ij}$.  These, in turn, may be related to the Yukawa couplings
$h_s$ (three real parameters in the diagonal basis) and $h_a$ (three complex
parameters).

Lepton family mixing may be
described by these three angles $\alpha^{ij}$ and three additional
angles $\beta^{ij}$ that diagonalize
the neutrino mass matrix.  For small mixing, the mixing angles for the
$W^-$, $Y^-$ and $Y^{--}$ charged currents are given by
$\alpha^{ij}-\beta^{ij}$, $\alpha^{ij}+\beta^{ij}$ and $2\alpha^{ij}$
respectively.  For massless neutrinos we are free to choose
$\beta^{ij}=\alpha^{ij}$ which ensures the $W^-$ charged current respects
lepton family.  In this case, family mixing is given by $2\alpha^{ij}$ for
both dilepton currents.

CP violation may occur in the gauge sector, but for massless neutrinos
would only show up in the off-diagonal dilepton couplings; whenever the CP
violating phase $\theta_{ij}$ shows up, $\alpha^{ij}$ must also be present.
Thus CP violation and lepton flavor violation are closely related, giving
the unusual prediction for the EDMs $d_\mu/d_\tau\approx-m_\tau/m_\mu$.
Additional CP violation may be present in the scalar sector, and need not be
related to lepton flavor violation.  The scalar contributions are only
important when $\alpha^{ij}\approx0$ and arise through a combination of a
three Higgs doublet model \cite{weinberg} and through the mixing of $T^{++}$
and $\eta^{++}$ \cite{zee}.

Experimentally, the non-observation of lepton flavor violation puts strong
restrictions on the mixing angles $\alpha^{ij}$.  The simplest way of
accommodating this is to postulate a discrete symmetry which prevents $\phi$
from coupling to the leptons, thus setting $h_a=0$.  This gives rise to a
purely symmetric mass matrix and vanishing $\alpha^{ij}$ (eliminating
dilepton mediated CP violation as well).

Since all leptons are embedded in a single $SU(3)_L$ representation, most
models of Majorana neutrino mass give rise to simple relations between
charged lepton and neutrino masses and mixing \cite{cubic}.  In particular,
when $h_a=0$ all mixing vanishes, $\alpha^{ij}=\beta^{ij}=0$, so the 331
model allows the interesting possibility of neutrino masses with no mixing.

Although our focus has been on the 331 model, the results are easily
generalized to encompass all models with dilepton gauge bosons resulting
from an $SU(3)$ generalization of the standard electroweak theory.  In
particular, the $SU(15)$ grand unified theory
\cite{su15fl,su15fk,su15fn,su15afn}
also leads to lepton flavor non-conservation via dilepton exchange.  This
point seems to have been missed in earlier analyses.

Similar to the 331 model, leptons in $SU(15)$ get symmetric and
anti-symmetric contributions to their mass matrices, this time from Higgs in
the $\bf120$ and $\bf105$ of $SU(15)$ respectively \cite{su15fk}.  Thus the
331 results for lepton masses and mixing, including CP violation governed by
dilepton exchange, are equally applicable to $SU(15)$ theory.  One crucial
difference, however, is that dileptons in $SU(15)$ may be very heavy, leading
to a natural suppression of rare lepton processes.  Indeed, much of the
appeal of the 331 model is that the new physics it predicts is guaranteed
to be below a few TeV, well within the reach of future colliders.  We look
forward to both direct and indirect tests that will soon conclusively decide
the fate of this model.

\section*{Acknowledgements}

J.T.L.\ would like to thank J.~Agrawal, P.~Frampton and P.~Krastev for
useful discussions.
This work was supported in part by the U.S.~Department of Energy under Grant
No.~DE-FG05-85ER-40219, the National Science Foundation under Grant
No.~PHY-916593 and by the Natural Science and Engineering Research Council
of Canada.

\appendix
\section*{Diagonalizing the charged lepton mass matrix}
In this appendix, we examine the new leptonic mixing matrix $V_Y$ and show how
it may be related to the lepton Yukawa couplings of Eq.~(\ref{eq:lepyuk}).
In particular we find a convenient way of determining the three angles and
three phases of $V_Y$ in terms of $h_s$ and $h_a$.

The unitary matrix $V_Y$ comes from diagonalization of the charged lepton
mass matrix, $M_\ell=(h_sv_3+h_av_1)/\sqrt{2}$ where $h_s$ ($h_a$) is
(anti-)symmetric.  In general, the VEVs $v_1$ and $v_3$ may be complex,
leading to CP violation in the scalar sector.  However, these phases can
always be absorbed into the Yukawa matrices.  Thus we assume this has
already been done, and that $v_1$ and $v_3$ are both real and
positive.

Starting with the lepton Yukawa interaction (\ref{eq:lepyuk}), we may
perform the $SU(3)_L$ invariant transformation in family space, $\psi'\to
U\psi'$ where $U$ is a unitary matrix.  This has no effect on the gauge
interactions, but replaces the Yukawa couplings $h_i$ by $Uh_iU^T$ in
(\ref{eq:lepyuk}).  Since $h_s$ is symmetric, we can always find a
matrix $U$ such that $Uh_sU^T$ is real and diagonal.  As a result, this
freedom allows us to pick $h_s={\rm diag}(h_s^1,h_s^2,h_s^3)$ where
$h_s^i$ is real and positive without any loss of generality.  This
immediately reduces the number of
parameters of $h_s$ from six complex entries to three real ones.

When $h_s$ is chosen in this form, $Uh_aU^T$ remains antisymmetric and has
three complex entries, $h_a^{12}$, $h_a^{13}$, and $h_a^{23}$.  In terms of
real parameters, this may be written as
$h_a^{ij}=y^{ij}e^{i\delta_{ij}}$, ($i\ne j$) where
$y^{ij}=-y^{ji}$ and $\delta^{ij}=\delta^{ji}$ .
In this special form, there are now only nine real Yukawa
parameters which are completely determined in terms of the three physical
charged lepton masses, three mixing angles and three physical phases of
$V_Y$.  In this way, the remaining nine real degrees of freedom present in a
general complex $3\times3$ mass matrix have been absorbed in the three
angles and six phases of the unobservable unitary matrix $U$.

Even with $h_s$ in this restricted form, diagonalization of $M_\ell$ is
non-trivial.  However, in order to suppress lepton flavor violation, it
is natural to assume that the antisymmetric contribution to $M_\ell$ is
small, $h_a|v_1|\ll h_s|v_3|$, so that the mass matrix is almost diagonal.  In
this limit, and to first order in $h_a$, we find the charged lepton masses
arise only from the symmetric Yukawa coupling, $m_i=h_s^i|v_3|/\sqrt{2}$.
The unitary matrices that
diagonalize $M_\ell$ are given by $E_L^{ij}\approx E_R^{ji}\approx
\delta^{ij}+\alpha^{ij}
e^{i\theta_{ij}}$ where the three angles $\alpha^{ij}=-\alpha^{ji}\ll1$ and
three phases $\theta_{ij}=-\theta_{ji}$ are given by
\begin{eqnarray}
\alpha^{ij}&\approx&{|v_1|\over |v_3|}
{y^{ij}\over (h_s^i)^2-(h_s^{\smash{j}})^2}
\sqrt{(h_s^i)^2+(h_s^{\smash{j}})^2-2h_s^ih_s^{\smash{j}}\cos2\delta_{ij}}
\nonumber\\
\tan\theta_{ij}&\approx&-{h_s^i+h_s^j\over h_s^i-h_s^{\smash{j}}}
\tan\delta_{ij}\ .
\end{eqnarray}
If all the Yukawa couplings are real ($\delta_{ij}=0$) so there is no
explicit CP violation, then the mixing angles have the
simple form $\alpha^{ij}\approx y^{ij}|v_1|/(h_s^i+h_s^j)|v_3|$.

Since $V_Y=E_R^TE_L^{\vphantom{T}}$, in this small mixing limit it has the form
$V_Y^{ij}=\delta^{ij} +2\alpha^{ij}e^{i\theta_{ij}}$ and is approximately
diagonal.  The physical picture that emerges is that the symmetric coupling,
$h_s$, gives rise to the charged lepton masses, whereas the antisymmetric
$h_a$ determines both lepton mixing and (CKM-like) leptonic CP violation.
An immediate consequence is that dilepton mediated lepton flavor violation
and CP violation are intimately related.  Both can be eliminated by
demanding $h_a=0$, as discussed in the main text.

\begin{figure}
\caption{The lepton flavor violating process $\mu\to3e$ via tree level
dilepton exchange.}
\label{fig:mu3e}
\end{figure}

\begin{figure}
\caption{The one-loop diagrams leading to $\mu\to e\gamma$.}
\label{fig:muegamma}
\end{figure}


\newpage
\input feynman
\bigphotons
\ \vskip3in
\begin{center}
\begin{picture}(40000,20000)
\THICKLINES
\drawline\fermion[\E\REG](3000,18000)[13000]
\drawarrow[\LDIR\ATTIP](\pmidx,\pmidy)
\global\advance\pfrontx by -2000
\global\advance\pfronty by -300
\put(\pfrontx,\pfronty){\large $\mu$}
\drawline\photon[\SE\FLIPPED](\pbackx,\pbacky)[13]
\global\advance\pmidx by -120
\global\advance\pmidy by -50
\drawarrow[\LDIR\ATBASE](\pmidx,\pmidy)
\global\advance\pmidx by -3500
\global\advance\pmidy by -1500
\put(\pmidx,\pmidy){\large $Y^{--}$}
\global\advance\photonlengthx by 13000
\drawline\fermion[\E\REG](\pfrontx,\pfronty)[\photonlengthx]
\drawarrow[\W\ATTIP](\pmidx,\pmidy)
\global\advance\pbackx by 1000
\global\advance\pbacky by -300
\put(\pbackx,\pbacky){\large $e$}
\drawline\fermion[\E\REG](\photonbackx,\photonbacky)[13000]
\drawarrow[\LDIR\ATTIP](\pmidx,\pmidy)
\global\advance\pbackx by 1000
\global\advance\pbacky by -300
\put(\pbackx,\pbacky){\large $e$}
\put(\pfrontx,\pfronty){\line(2,-1){13000}}
\global\advance\pfrontx by 6500\global\advance\pfronty by -3250
\put(\pfrontx,\pfronty){\vector(2,-1){0}}
\global\advance\pfrontx by 6500\global\advance\pfronty by -3250
\global\advance\pfrontx by 1000
\global\advance\pfronty by -300
\put(\pfrontx,\pfronty){\large $e$}
%
%
\end{picture}
\end{center}
\vskip 1in
\centerline{Fig.~1}
\newpage
%
%
\ \vskip.2in
\begin{center}
\begin{picture}(40000,25000)
\THICKLINES
\drawline\fermion[\E\REG](2370,22000)[10000]
\drawarrow[\LDIR\ATTIP](\pmidx,\pmidy)
\global\advance\pfrontx by -2000
\global\advance\pfronty by -300
\put(\pfrontx,\pfronty){\large $\mu$}
\drawline\photon[\E\REG](\pbackx,\pbacky)[15]
\global\advance\pmidy by -160
\global\advance\pmidx by -20
\drawarrow[\LDIR\ATBASE](\pmidx,\pmidy)
\global\advance\pmidy by 1150
\global\advance\pmidx by -1300
\put(\pmidx,\pmidy){\large $Y^{--}$}
\drawline\fermion[\E\REG](\pbackx,\pbacky)[10000]
\drawarrow[\LDIR\ATTIP](\pmidx,\pmidy)
\global\advance\pbackx by 1000
\global\advance\pbacky by -300
\put(\pbackx,\pbacky){\large $e$}
%
\multroothalf\photonlengthx
\drawline\fermion[\SW\REG](\photonbackx,\photonbacky)[\photonlengthx]
\drawarrow[\LDIR\ATTIP](\pmidx,\pmidy)
\global\advance\pmidy by -1000
\global\advance\pmidx by 500
\put(\pmidx,\pmidy){\large $e$, $\mu$, $\tau$}
\drawline\fermion[\NW\REG](\pbackx,\pbacky)[\photonlengthx]
\drawarrow[\LDIR\ATTIP](\pmidx,\pmidy)
\drawline\photon[\S\REG](\pfrontx,\pfronty)[10]
\global\advance\pbacky by -2000
\global\advance\pbackx by -200
\put(\pbackx,\pbacky){\large $\gamma$}
%
\end{picture}
\end{center}
%
%
\vskip.5in
\begin{center}
\begin{picture}(40000,25000)
\THICKLINES
\drawline\fermion[\E\REG](3000,22000)[10000]
\drawarrow[\LDIR\ATTIP](\pmidx,\pmidy)
\global\advance\pfrontx by -2000
\global\advance\pfronty by -300
\put(\pfrontx,\pfronty){\large $\mu$}
\drawline\photon[\SE\FLIPPED](\pbackx,\pbacky)[11]
\global\advance\pmidx by 50
\global\advance\pmidy by 120
\drawarrow[\LDIR\ATBASE](\pmidx,\pmidy)
\drawline\photon[\NE\FLIPPED](\pbackx,\pbacky)[11]
\global\advance\pmidx by -120
\global\advance\pmidy by 50
\drawarrow[\LDIR\ATBASE](\pmidx,\pmidy)
\global\advance\pmidy by -1500
\global\advance\pmidx by 500
\put(\pmidx,\pmidy){\large $Y^{--}$}
\double\photonlengthx
\drawline\fermion[\W\REG](\pbackx,\pbacky)[\photonlengthx]
\drawarrow[\LDIR\ATTIP](\pmidx,\pmidy)
\global\advance\pmidy by 1000
\global\advance\pmidx by -1700
\put(\pmidx,\pmidy){\large $e$, $\mu$, $\tau$}
\drawline\fermion[\E\REG](\pfrontx,\pfronty)[10000]
\drawarrow[\LDIR\ATTIP](\pmidx,\pmidy)
\global\advance\pbackx by 1000
\global\advance\pbacky by -300
\put(\pbackx,\pbacky){\large $e$}
%
\drawline\photon[\S\REG](\photonfrontx,\photonfronty)[10]
\global\advance\pbacky by -2000
\global\advance\pbackx by -200
\put(\pbackx,\pbacky){\large $\gamma$}
%
\end{picture}
\end{center}
\vskip .3in
\centerline{Fig.~2}

\begin{references}
\bibitem{fn:triplet} We could equally well have used triplets \cite{pleitez},
which would lead to a different embedding of the electric charge, but gives an
otherwise identical theory.
\bibitem{fn:alt} Several alternatives have been proposed where the sextet
Higgs is not present \cite{duong,ppt}.  However, as they involve either
adding new particles \cite{duong} or breaking lepton number \cite{ppt},
they may be considered non-minimal versions of the 331 model.
\bibitem{fn:Tvev} Another possibility is to break lepton number with
$\langle T\rangle\ne0$.  In order to avoid a massless triplet Majoron, this
necessitates introducing explicit lepton number violation in the scalar
potential.
\bibitem{fn:notzee} In the model of Ref.~\cite{zee}, the triplet and singlet
have unrelated Yukawa couplings.  Here, both couplings are identical so
$T^{++}$--$\eta^{++}$ CP violation is somewhat more constrained.
\end{references}
\end{document}